\documentclass[conference]{IEEEtran}
\usepackage{amsmath,amssymb,enumerate}
\usepackage{cite,graphics,graphicx}
\usepackage{graphicx}
\usepackage{subfigure}
\usepackage{makeidx}
\usepackage{hyperref}
\usepackage{algorithm}
\usepackage{algorithmic}

\usepackage{multirow}
\usepackage{amsmath}
\usepackage{amssymb}
\usepackage{xcolor}
\usepackage{epsfig}
\usepackage{graphics}
\usepackage{listings}

\ifCLASSINFOpdf
\else
\fi
\begin{document}
%
\title{Massively Parallel Ray Tracing Algorithm Using GPU}


\author{\IEEEauthorblockN{Yutong Qin\IEEEauthorrefmark{1}\IEEEauthorrefmark{3},
Jianbiao Lin\IEEEauthorrefmark{2}\IEEEauthorrefmark{3},
Xiang Huang\IEEEauthorrefmark{1}\IEEEauthorrefmark{4}}
\IEEEauthorblockA{Department of Computer Science and Engineering\\
Sichuan Unversity, Jinjiang College, Pengshan, Sichuan, 620860, China}

\IEEEauthorblockA{\IEEEauthorrefmark{1}1359191149@qq.com
\IEEEauthorrefmark{2}zeroyuebai@hotmail.com}
}

%


\maketitle

\begin{abstract}
Ray tracing is a technique for generating an image by tracing the path of light through pixels in an image plane and simulating the effects of high-quality global illumination at a heavy computational cost. Because of the high computation complexity, it can't reach the requirement of real-time rendering. The emergence of many-core architectures, makes it possible to reduce significantly the running time of ray tracing algorithm by employing the powerful ability of floating point computation. In this paper, a new GPU implementation and optimization of the ray tracing to accelerate the rendering process is presented.
\end{abstract}


%
\IEEEpeerreviewmaketitle

\section{Introduction}
Photorealistic rendering is an rendering process of the reflection effects of real shadow rays. Unlike the pipeline of real-time rendering, it requires to achieve the high quality of reality to guarantee its authenticity is hard to verify, thus realistic illumination and materials need quite complicated and accurate simulation. Physics-based rendering technology can achieve photo-realistic rendering, but the huge computational cost makes real-time photorealistic rendering of a image can not be generated in time. On the contrary of the pipeline rendering, the former sacrifices reality for high-speed rendering and real-time performance. The latter attenuates high-speed rendering and real-time performance to  dramatically enhance the effect of reality. Because of these properties of ray tracing, it has been widely applied in film, advertising, animation, and other visual industries.
Ray tracing is other than the widely used technique in interactive computer graphics, rasterization. Based on physical optics theorem, ray tracing can simulate the light propagation in the real world and calculate the distribution of radiation. Because of the heavy computational complexity of simulating the light propagation, rendering an image usually takes tens of minutes to several hours, so to product the high-quality real images, we generally requires specialized high-performance equipment. Before the GPU computing was proposed, ray tracing technique has always been a very time consuming work.

In recent years, the emergence of parallel computing based on GPU architectures,
many researchers are interested in employing the powerful ability of floating point
computation to improve the efficiency of ray tracing algorithm because of the low entry threshold. Unlike the design philosophy of CPU architecture, GPU is generally comprised of hundreds of thousands of stream processors. Many-core architecture is split into a large number of much smaller cores and each core is an in-order, heavily multi-threaded, single-instruction issue processor that shares its control and instruction cache with other cores. So data-intensive applications can easily harness the potential power of GPUs. Because there are a large number of calculation in ray tracing algorithm, for example, traverse, circulation and intersection, all of these calculation can be decomposed into
independent subtasks to execute in parallel. It is not difficult to imagine how the ray tracing's performance varies under GPU architecute.

In modern software processes, the program sections often exhibit a rich amount of data parallelism, a property that allows many arithmetic operations to be performed on program data structures in a simultaneous manner. CUDA devices accelerate the execution of these applications by obtaining a large amount of data parallelism. Besides CUDA, several tools including language, library, and compiler directives are still used. For example, OpenCL, which is a framework for writing programs, can be executed across heterogeneous platforms consisting of CPUs, GPUs, digital signal processors (DSPs), and other processors. Considering good characteristics of OpenCL, such as flexibility, portability, versatility, we used OpenCL to optimize and accelerate ray tracing algorithm.

\section{The problem}
Since the vast majority of ray tracing applications today perform on CPU architecure, it makes the efficiency of ray tracing have direct relation with Cycles Per Instruction (CPI) and cycle rate.
CPI is determined by Instruction Set Architecture (ISA). Because of the bottleneck of Moore's law,
CPU manufacturers have gradually reached the limit of clock frequency. Thus, serial program can not essentially improve the efficiency of ray tracing. However,
 today it has not taken a gigantic leap forward even in multi-core CPU architecture.

 To solve these problems, many researchers designed lots of the acceleration of ray tracing algorithm, including space partition, bounding box, spatial sorting, and so forth. Because these methods exclude those objects and lights who do not involve in ray tracing, the optimized scene do greatly reduce the time overhead of ray tracing. But, more or less, every optimization method has limitations. For example, space partition's efficiency is generally limited by intensive scenes.

 On the other hand, there are hardwares specifically designed for ray tracing. For example, light processing unit developed by Stanford, but poor universality, only a few people can use these dedicated hardwares. Another solution is distributed computing using cluster. It splits the problem into independent subproblem and these tasks will be mapped into the different computer nodes. The cost of that is significant, in the meantime, it's extremely hard to guarantee load balancing.

 It is becoming increasingly common to use a general purpose graphics processing unit as a modified form of stream processor. This concept turns the massive computational power of a modern graphics accelerator's shader pipeline into general-purpose computing power. GPU can be used for many types of embarrassingly parallel tasks including ray tracing. They are generally suited to high-throughput type computations that exhibit data-parallelism to exploit the wide vector width SIMD architecture of the GPU.

In general, GPU allows to launch tens of thousands of lightweight threads to execute the same kernel function simultaneously. with this feature, independent lightweight threads can take the place of multi-level iterations and massively parallel ray tracing algorithm. So GPU can greatly improve the efficiency of ray tracing.

\section{Ray tracing}
In computer graphics, ray tracing is a technique for generating an image by tracing the path of light through pixels in an image plane and simulating the effects of its encounters with virtual objects. If the ray intersects with some objects, according to the theorem of radiosity, the color value of the related point in the image plane can be calculated by this method using some parameters, for example, materials, normal vector at the intersection point, light distribution, and so on. More specifically, to get the color value at one point, it is a critical part to calculate the radiance of the opposite direction of the ray casting at this point.

\begin{figure}[H]
\centering
\includegraphics[width=0.40\textwidth]{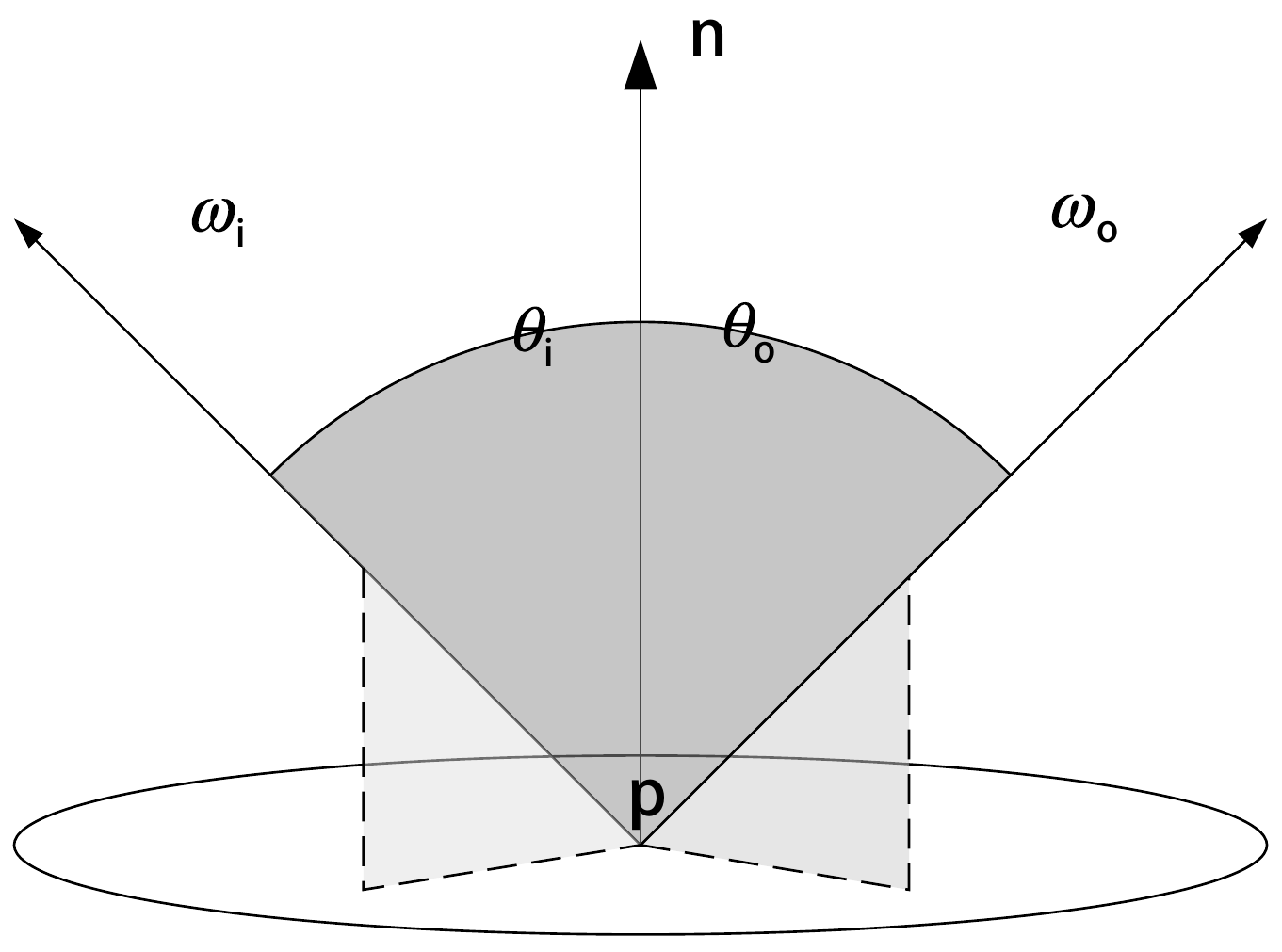}
\caption{In the radiosity model, $\textit{$w_{i}$}$ and $\textit{$w_{o}$}$ represent the directions of incident light and emergent light. }
\label{fig1}
\end{figure}

As shown in Fig. \ref{fig1}, point $\textit{p}$ is an random point on the object surface. It's the origin of an eclipse and that eclipse is the integration region of point $\textit{p}$. By convention, $\textit{$w_{i}$}$ points to light source or one sampling point on its surface, and $\textit{$w_{o}$}$ can finally reach the viewing plane.
Set the radiance along the reflect light of $\textit{$w_{o}$}$ to $L$. Through the Lambert's emission law, the equation is derived as follows:

\begin{equation}\label{e1}
  L = \frac{d^{2}\Phi}{dA\cos{\theta}dw}
\end{equation}

As Eq. (\ref{e1}) shows, $d^{2}\Phi$ means the radiation power which emits from the surface element $dA$ to the solid angle $dw$. Through the formula of irradiance:

\begin{equation}\label{e2}
  E = \frac{d\Phi}{dA}
\end{equation}

In considering of the premise of incident direction, Eq. (\ref{e1}) is substituted into Eq. (\ref{e2}) as follows:

\begin{equation}\label{e3}
  dE_{i}(p,w_{i}) = L_{i}(p,w_{i})\cos{\theta}dw_{i}
\end{equation}

In Eq. (\ref{e3}), the received irradiance $dE_{i}(p,w_{i})$ at the point $p$ can be calculated by the radiance
$L_{i}(p,w_{i})$ at that point. Obviously, incident angle $\theta$ is the other impact factor to the final result.
For general materials, irradiance is proportional to radiance, that is, with greater radiosity, comes greater reflection of radiosity at the same point. Thus, the following relation holds certainly:

\begin{equation}\label{e4}
  dL_{o}(p,w_{o}) \propto dE_{i}(p,w_{i})
\end{equation}

If bidirectional reflectance distribution function (BRDF) is used to define the scale factor, Eq. (\ref{e4} can be transformed as Eq. (\ref{e5}):

\begin{equation}\label{e5}
  dL_{o}(p,w_{o}) = f_{r}(p,w_{i},w_{o})dE_{i}(p,w_{i})
\end{equation}

And then, Eq. (\ref{e3}) is substituted into Eq. (\ref{e5}) as following:

\begin{equation}\label{e6}
  dL_{o}(p,w_{o}) = f_{r}(p,w_{i},w_{o})L_{i}(p,w_{i})\cos{\theta}dw_{i}
\end{equation}

If the surface of object is self illuminated material, besides the reflection of radiosity, the surface emits radiance also include it emits radiosity by itself. Set self illuminated material emits radiance to $L_{e}$. $L_{e}$ is added into Eq. (\ref{e6}) as below:

\begin{equation}\label{e7}
  dL_{o}(p,w_{o}) = L_{e} + f_{r}(p,w_{i},w_{o})L_{i}(p,w_{i})\cos{\theta}dw_{i}
\end{equation}

As shown in Fig. \ref{fig1}, assume that consider only the single incident direction $w_{i}$, Eq. {\ref{e7}} can calculate the integration of radiance in any directions. However, it's impossible that the irradiance at point $p$ simply originate from single direction. In reality, point $p$ would receive irradiance of all directions in the hemisphere region above that point. Radiance is obtained by integration of Eq. (\ref{e7}) as follows:

\begin{equation}\label{e8}
  L_{o}(p,w_{o}) = L_{e} + \int_{2\pi^{+}} f_{r}(p,w_{i},w_{o})L_{i}(p,w_{i})\cos{\theta}dw_{i}
\end{equation}

\begin{figure}[H]
\centering
\includegraphics[width=0.42\textwidth]{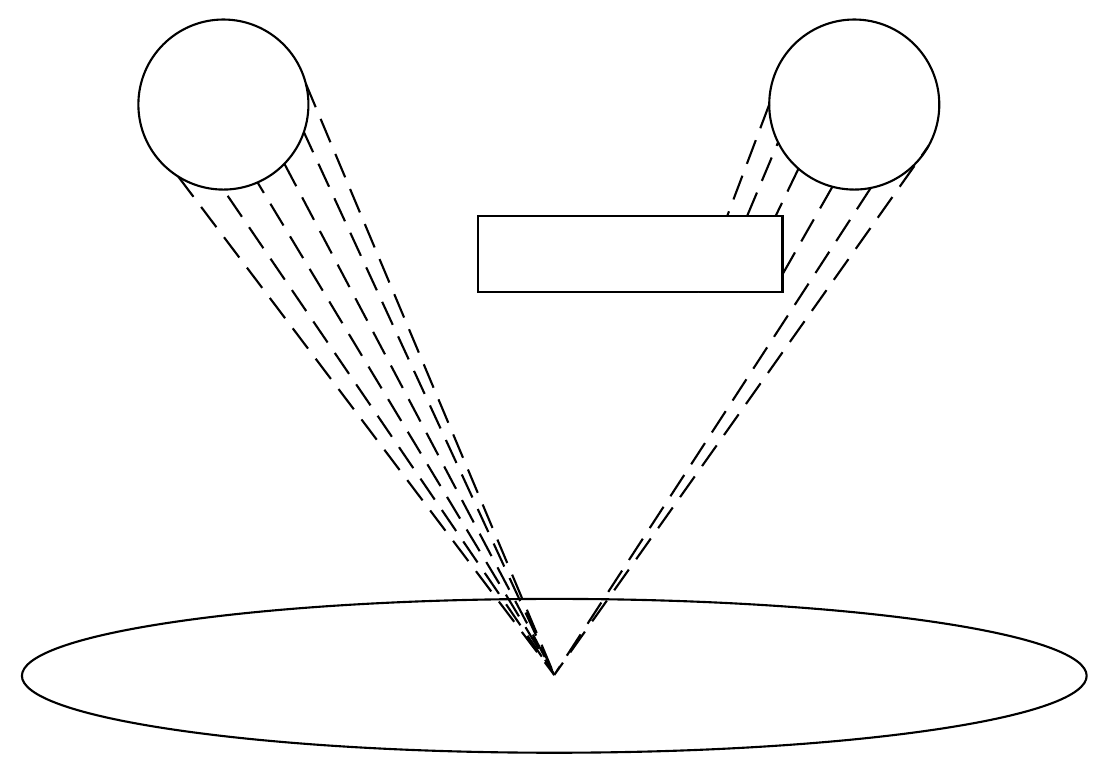}
\caption{In local illumination model, source lights all have radiosity effects on point $p$.}
\label{fig2}
\end{figure}

Although Eq. (\ref{e8}) provides the equation to calculate the whole radiance in the surface of objects, apparently it can't be solved for straight away. There are a couple of reasons for this. First, Eq. (\ref{e8}) contains a constant integral limitation which can be seen as Fredholm integral equation of the second kind. Second, because computer can not precisely simulate irradiance of all directions in the hemisphere region. Even in the global illumination model, it is unable to trace all the lights at one point of object's surface. Thus, the mathematical model described in Eq. (\ref{e8}) should be simplified.  We can recursively trace a small amount of indirect reflected light on  object's surface. Recursion depth depends on the number of light reflection. So the majority of integral calculation is concentrated on radiosity of sampling points on the surface of light source, as shown in Fig. \ref{fig2}.

In Fig. \ref{fig2}, to get the radiance along light to viewing plane at point $p$, calculating the   received irradiance of that point using Eq. (\ref{e8}) is necessary. Point $p$ can receive the whole radiosity from light source no.1 and partial that from no.2. The process of integration need to traverse all sampling points on the surface of both regions and determine one by one whether the light is obstructed by objects. For example, the object in Fig. \ref{fig2} blocked some radiosity from light source no.2. The blocked radiation did not make a contribution to the lighting of point $p$ at all. Afterwards, integrating the received radiosity at point $p$.  This process is generally the most time-consuming part of ray tracing which depends on the number of light sources and geometries, the intersection complexity of geometries, the number of sampling points on the surface of light source and so on. If the process of rendering using anti-aliasing technology, each pixel will cast more light and finally the pixel will take the average value of these colors.
The pseudocode of local illumination ray tracing can be depicted as follows:

\begin{algorithm}[h]
\caption{Local Illumination Ray Rracing}
\begin{algorithmic}[1]

\FOR{each light of each pixel in the scene}

\FOR{each object in the scene}

\FOR{one light intersects with one object}

\FOR{each sampling point of each source light}

\STATE emit a shadow light $r$ from point $p$ to that sampling point

\FOR{each object in the scene}

\IF{$r$ intersects with one object}
\STATE \textbf{break}
\ELSE
\STATE calculate the irradiance using Eq. (\ref{e8})
\ENDIF

\STATE accumulate all the received irradiance
\ENDFOR

\ENDFOR

\ENDFOR
\STATE accumulate the color value of each light
\ENDFOR
\STATE take the average value of these pixel's colors
\ENDFOR

\label{alg1}
\end{algorithmic}
\end{algorithm}

As shown in Alg. \ref{alg1},  multilevel nest iterations exhibit a rich amount of data parallelism. The pseudocode only considers the radiosity point p received directly. In the global illumination, besides radiosity from source light, it also includes reflection radiosity from objects, so the program need to be modified as an recursive version. However, the performance of serial execution is inefficient.

\section{Paralel optimization}

\subsection{parallel ray tracing}

\begin{figure}[H]
\centering
\includegraphics[width=0.42\textwidth]{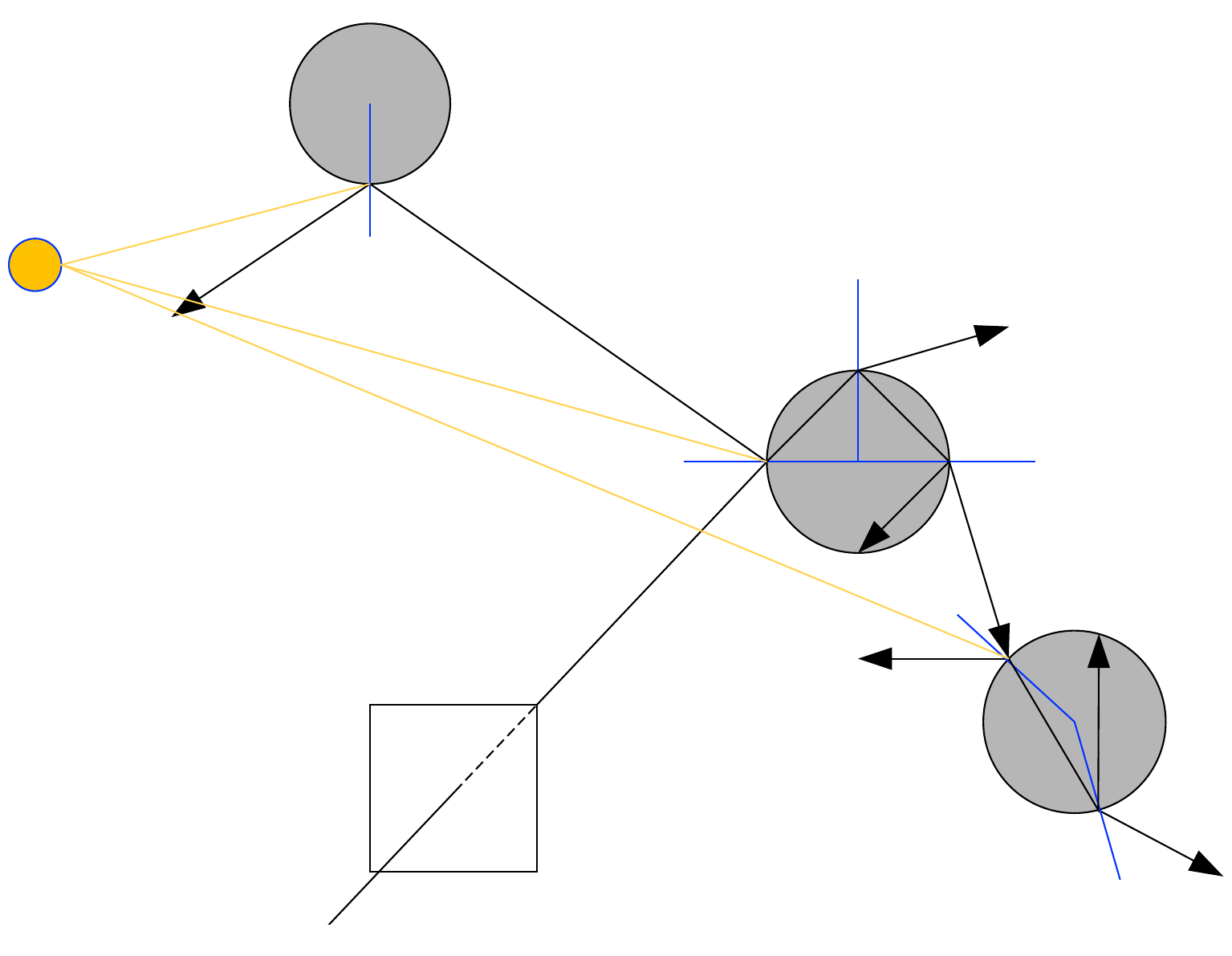}
\caption{Lights occurr radiosity on the other objects through reflection and refraction}
\label{fig3}
\end{figure}

In traditional global illumination model, when a single light intersects with object in the scene, it will produce some of secondary lights. Some secondary are shadow lights which can be used to check the visibility of light sources. Besides that, all the others are treated as new generation lights to spread again (intersection test and radiosity calculation), as shown in Fig. \ref{fig3}.

Recursion method is used to trace secondary lights until they reach the maximum recursion depth. Secondary lights
occurr radiosity on the other objects, so global illumination is also called indirect illumination.

Since OpenCL kernel don't support the property of recursion, recursion need to be transformed into iterations and the number of iterations is used to simulate the recursion depth. When a single light reaches a point on the surface of one object, derived shadow lights at intersection only need to sample every light source once. They traverse all the sampling points of each light source is unnecessary. when all the lights recursively sample the surface of light source just once, the process of rendering will be suspended and the image will be updated. The next ray tracing will select another sampling point and start the same work at once. Then overlapping new color value onto the pixel. Iterations are to simulate the integration of the radiosity of sampling points on the light source's surface.

\begin{figure}[H]
\centering
\includegraphics[width=0.48\textwidth]{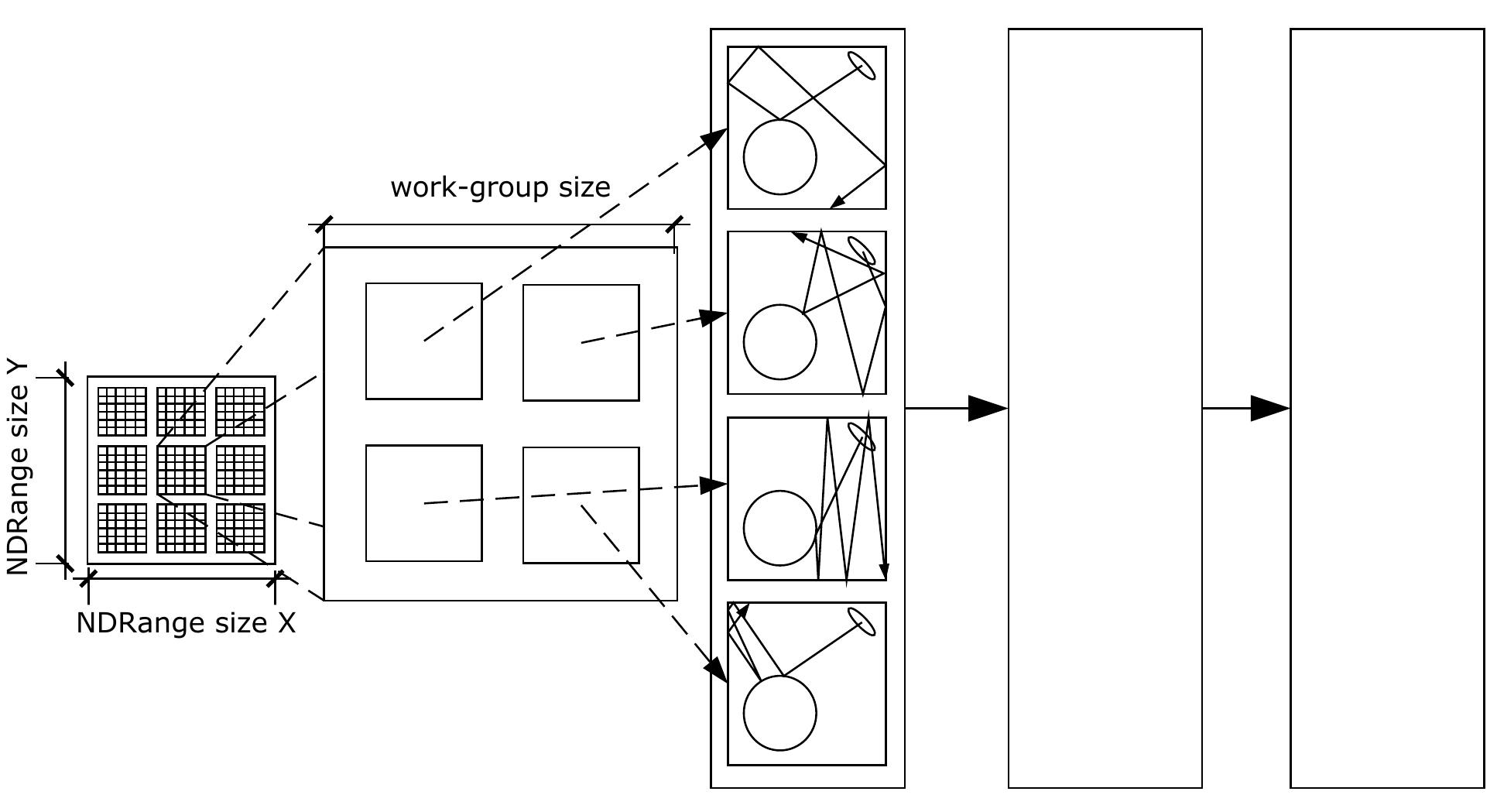}
\caption{Overview of parallel ray tracing algorithm using GPU}
\label{fig4}
\end{figure}

In Fig. \ref{fig4}, under GPU architecute, each kernel thread traces a single light and  it can obtain the final color value of the light. When all threads execute the kernel function once, the intermediate value will be added into the pixels. To render a image, the same kernel function should be launched iteratively.

\subsection{GPU Kernel Function}

To simplify the programming model, this paper only study the rendering of sphere. The implicit equation of sphere can be represented in vectorial form.

\begin{equation}\label{e9}
  (p - c)\cdot(p - c) - r^{2} = 0
\end{equation}

Linear equation can be expressed as below:

\begin{equation}\label{e10}
  o + td
\end{equation}

Eq. (\ref{e10}) is substituted into Eq. (\ref{e9}) as following:

\begin{equation}\label{e11}
  (d\cdot d)t^{2} + 2[(o - c)\cdot d]t + (o - c)\cdot(o - c) - r^{2} = 0
\end{equation}

Eq. (\ref{e11}) can be regarded as a quadratic equation. So t is a dependent variable, the formula
can be transformed as follows:

\begin{equation}\label{e12}
  t = \frac{-b \pm \sqrt{b^{2} - 4ac}}{2a}
\end{equation}

Note that the variables $a$, $b$ and $c$ can be calculated as below: $a = d\cdot d$,  $b = 2(o - c)\cdot d$ and  $c = (o - c)\cdot(o - c) - r^{2}$.

 Eq. \ref{e12} determines whether a single light intersects with sphere. If so, the coordinate of intersection can be calculated. To calculate the process of intersection more efficiently, We need to transform the equation into OpenCL kernel function. In combination with Eq. (\ref{e8}), massively parallel integration can achieve the goal of improving the efficiency of rendering.

\section{Results and Discussion}

Tests were conducted on a system composed of an Intel Core i7-2720QM CPU running at 2.20GHz, with 1600MHz and 4GB DDR3 DRAM. This platform also had a ATI Radeon HD 6750M GPU. The scene file provided by David Bucciarelli and the scene resolution is 640 $\times$ 480.

\begin{figure}[H]
\centering
\includegraphics[width=0.37\textwidth]{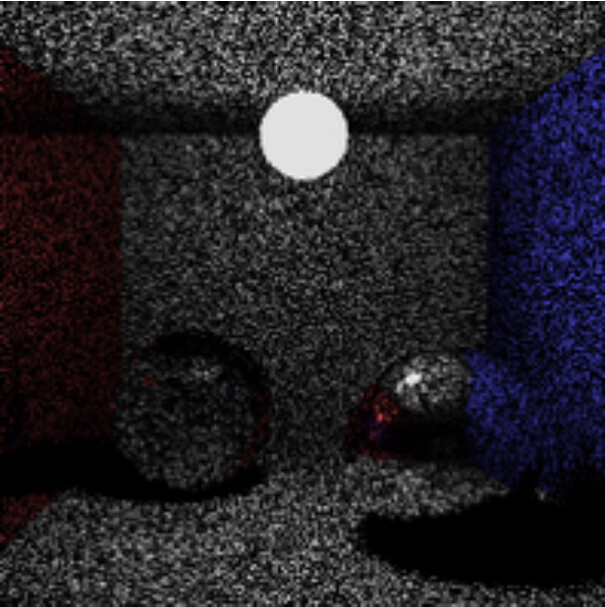}
\caption{The first rendering took only 0.508 seconds to generate the image.}
\label{fig5}
\end{figure}

\begin{figure}
\centering
\includegraphics[width=0.37\textwidth]{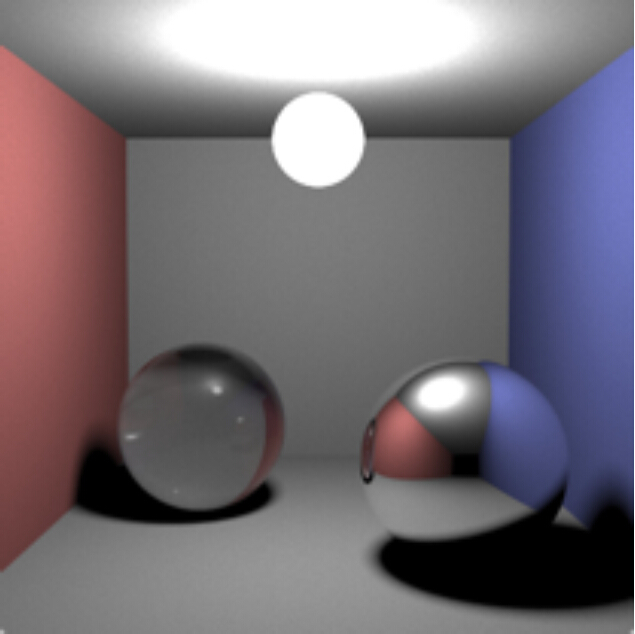}
\caption{After 6 seconds, the image showed better rendering effects.}
\label{fig6}
\end{figure}

In Fig. \ref{fig5}, the image was generated using local illumination model while a single cycle of rendering was finished. Since parallel rendering once only selects one sampling point, partial region of the image produced amounts of black noise. When more and more cycles are completed, sampling points will cover most of the pixels in the scene, thus, the image will show better rendering effects (see Fig. \ref{fig6}). As time goes on, more sampling points will be rendered, the image will become more accurate.

\begin{figure}
\centering
\includegraphics[width=0.37\textwidth]{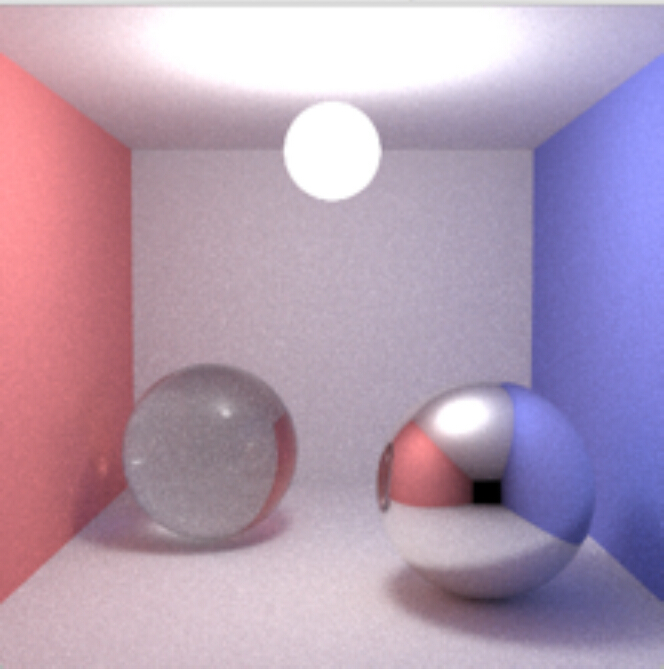}
\caption{Overview of radiosity model, }
\label{fig7}
\end{figure}

\begin{figure}
\centering
\includegraphics[width=0.5\textwidth]{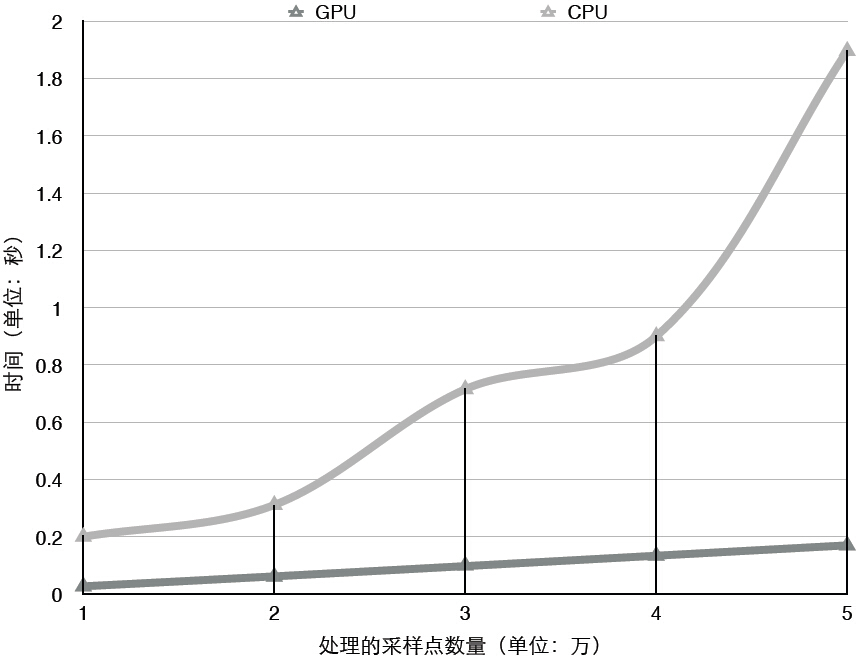}
\caption{Overview of radiosity model, }
\label{fig8}
\end{figure}

In Fig. \ref{fig7}, the image was generated using global illumination model in the same scene. Its recursion depth was 6 and it took 20 seconds to generate this image. The experimental result shows that parallel ray tracing based on GPU significantly improves rendering effects. Here, as shown in Fig. \ref{fig8}, a comparative evaluation of ray tracing to process the same number of sampling points under two different platforms, multi-core (i7-2720QM CPU) and many-core (ATI Radeon HD 6750M GPU) is proposed.

\section{Conclusion}

\section*{Acknowledgment}

This work was completely supported by  the department of Civil Engineering and the department of Computer Science$\&$Engineering Jinjiang College, Sichuan University. All authors read and approved the final manuscript. We are deeply indebted to somebody for their encouragement and support.




\begin{thebibliography}{11}


\bibitem{suffern2007ray}
Suffern, Kevin Geoffrey and Kevin Suffern, \emph{Ray Tracing from the Ground up}, AK Peters, 2007.

\bibitem{Physically}
Pharr, Matt and Greg Humphreys, \emph{Physically based rendering: From theory to implementation}, Morgan Kaufmann, 2010.



\bibitem{Hongning}
Weiping, Li Hongning Feng Jie Yang, and Bai Fengxiang, \emph{Spectral-Based Rendering Method and Its Application in Multispectral Color Reproduction}, Laser \& Optoelectronics Progress 12 (2010): 020.

\bibitem{Christen}
Christen, Martin, \emph{Ray tracing on GPU}, Master's thesis, Univ. of Applied Sciences Basel (FHBB), Jan 19 (2005).

\bibitem{Tomas}
Akenine-Moller, Tomas, Eric Haines and Naty Hoffman, \emph{Real-time rendering}, AK, 2002.

\bibitem{Macbeth}
Macbeth color checker patches data, Munsell Color Science Laboratory, http://www.cis.rit.edu/research/mcsl2/online/cie.php

\bibitem{OpenCL}
Munshi, Aaftab, Benedict Gaster, Timothy G. Mattson and Dan Ginsburg, \emph{OpenCL programming guide}, Pearson Education, 2011.

\bibitem
@article{dutre2003global,
  title={Global illumination compendium},
  author={Dutr{\'e}, Philip},
  journal={Computer Graphics, Department of Computer Science Katholieke Universiteit Leuven},
  year={2003}
}

\bibitem
@inproceedings{kajiya1986rendering,
  title={The rendering equation},
  author={Kajiya, James T},
  booktitle={ACM Siggraph Computer Graphics},
  volume={20},
  number={4},
  pages={143--150},
  year={1986},
  organization={ACM}
}
\bibitem
@book{van2014computer,
  title={Computer graphics: principles and practice},
  author={Van Dam, Andries and Feiner, Steven K},
  year={2014},
  publisher={Pearson Education}
}

\end{thebibliography}
%

\end{document}